\documentclass[10pt,a4paper]{article}
\usepackage{jheppub_kim}
\usepackage{pdflscape}
\usepackage{amsmath}
\usepackage{amssymb}
\usepackage{dcolumn}
\usepackage{bm}
\usepackage{color}
\usepackage{epsfig}
\usepackage{amsfonts}
\usepackage{graphicx}
\usepackage{subfigure}
\usepackage{dcolumn}
\UseRawInputEncoding

\begin{document}

\title{\Large{Gravitational waves driven by Holographic dark energy}}

\author[a]{Sayani Maity,}
\author[b]{Prabir Rudra}

\affiliation[a]{Department of Mathematics, Sister Nivedita University, DG-1/2, Action Area 1, New Town, Kolkata-700 156, India}

\affiliation[b] {Department of Mathematics, Asutosh College,
Kolkata-700 026, India.}

\emailAdd{sayani.maity88@gmail.com}
\emailAdd{prudra.math@gmail.com}


\abstract{In this paper, we have studied the effects of holographic dark energy on the evolution of gravitational waves. The background evolution of gravitational waves in a flat FRW universe is considered and studied in the presence of various holographic dark energy models. The perturbation equations governing the evolution of the gravitational waves have been constructed and solutions are obtained. These solutions are studied in detail to get a proper understanding of the characteristics of the gravitational waves in the presence of holographic dark energy. The work can be a significant tool in studying different dark energy models comparatively using the features of the gravitational wave evolution.}

\maketitle

\section{Introduction}
The universe' late-time acceleration has been a hotly debated topic over the last twenty years. Large-scale structure observations \cite{cmb2, cmb3, cmb4}, Cosmic Microwave Background Radiation \cite{cmb1}, and type Ia supernovae \cite{sn1, sn2, sn3}, among other recent developments in observational cosmology, validate this property of the universe. The theory being developed to explain this universe-processing apparatus, however, is what really has to be addressed. It has been widely accepted for more than a century that the most successful fundamental gravitational theory for understanding the large-scale structure of the universe is general relativity (GR). In cosmology, the matter source and the FLRW spacetime give precise answers for the scale factor $a(t)$, which aids in our understanding of the universe's expansion. We can better understand the expansion of the universe by using the exact solutions for the scale factor $a(t)$ that the matter source and the FLRW spacetime in cosmology provide. To recover FLRW spacetime, the cosmos must undergo two distinct phases of accelerated expansion. The dynamics of early time inflation is one such instance, which can be investigated by adding a scalar field to the Einstein-Hilbert action. Moreover, the current rapid expansion can be explained by adding a cosmological constant to the Einstein field equation. This relatively simple model, termed as the $\Lambda$CDM model, fits all the available observations and explains the universe's late-time cosmic acceleration. Regrettably, vacuum energy scale fine-tuning issues plague this lovely model \cite{vc1, vc2}. Therefore, it is essential to take into account different strategies or generalizations of basic theories of gravity.

Significant progress in the study of gravitational physics was made in 2015 when the LIGO collaboration directly detected the gravitational wave (GW) signal from the binary black hole coalescence GW150914 \cite{gw11}. Later in 2017, a GW signal was detected by the LIGO-Virgo detector network coming from the merging binary neutron star GW170817 \cite{gw12}. This signal was linked to a brief $\gamma$-ray burst, GRB170817A \cite{gw13}, that was picked up by INTEGRAL \cite{gw18, gw19} and the Fermi Gamma-Ray Burst Monitor \cite{gw14, gw15, gw16, gw17}. The exciting discovery of the GW event and its electromagnetic (EM) equivalent by coincidence marked the beginning of the multi-messenger era in astrophysics \cite{gw20}.

Theoretically gravitational waves are an amazing consequence of Einstein's general theory of relativity that has captivated scientists and astronomers alike. Crucial data regarding the amplitude of these spacetime ripples can be obtained by measurements of polarization and cosmic microwave temperature anisotropy \cite{Crittenden1, Crittenden2}. Given the behavior of various types of waves observed in nature, it makes sense that perturbations in the gravitational field may propagate like waves. When Einstein demonstrated that gravitational radiation, or gravitational waves, is a natural byproduct of his theory, the connection between gravitational waves and the fundamental concepts of general relativity was further established. In the limit of small deviations from Euclidean space-time (or Minkowski space), Einstein's field equations yield a linear wave equation with plane wave solutions. These solutions describe transverse metric disturbances of Minkowski space that travel at the speed of light and have characteristics akin to electromagnetic waves. While electromagnetic and gravitational waves have many similarities, it's vital to keep in mind that they differ as well, thus drawing conclusions from this comparison should be done carefully. When gravitons decoupled in the early Universe, gravitational waves broke away from matter. Therefore, in order to restrict and differentiate cosmological parameters in various cosmological models, these gravitational waves play a crucial role. It's interesting to note that primordial gravitational waves originated from vacuum fluctuations. The matter content of an imaginary early Universe can be described by a cosmological model that aims to unify dark energy and dark matter, the two dark sectors of the Universe. This is where a smooth transition takes place between an early de Sitter-like phase and a radiation-dominated era. Gravitational waves can provide insight into the times when the underlying cosmic dynamics fluctuated, given all the potential dark energy hypotheses in cosmology. Consequently, several investigations of late have focused on gravitational waves \cite{Buonanno97, Gasperini1, Fabris1, Infante, Riazuelo, Fabris2, Santos, Lopez, Zhang, Debnath1, Debnath2}.

The holographic principle, originating from black hole thermodynamics \cite{14, 15}, states that a system's area, not its volume, determines its entropy. Holographic dark energy (HDE) \cite{16, 17, 18} is a result of this holographic idea, which also has connections to string theory \cite{14, 19}. A quantum field theory \cite{20} is generally linked to an ultraviolet cutoff, which is the maximum distance permitted by the framework. This ultraviolet cutoff is intimately linked to vacuum energy, a form of holographic dark energy. For a thorough summary of HDE, the reader may refer to \cite{21}. Numerous studies have examined HDE in both its basic and enlarged variants and the model has consistently shown to be highly effective \cite{22, 23, 24, 25, 26, 27, 28}. One of the key success factors for HDE models has been their compatibility with observational data \cite{29, 30, 31}. It should be noted that the HDE model differs greatly from the other models of dark energy that are based on cosmic fluids or scalar-tensor theory. The early inflation has also been explained by the holographic concept in the literature \cite{in1, in2, in3}. Because of the small size of the early Universe, the holographic energy density played a crucial role in inducing inflation. It was also discovered that this inflation can be matched with the 2018 Planck measurements.

Herein, our primary objective is to explore the characteristics of gravitational waves within the context of holographic dark energy in a Friedmann-Robertson-Walker (FRW) spacetime. In order to achieve this, we study how the dynamics of gravitational waves are affected by the presence of holographic dark energy. We investigate the gravitational wave evolution in a flat FRW universe taking different models of HDE into consideration. Treating gravitational waves has the benefit of being highly sensible to scale factor behavior, even when the matter content is not immediately visible. Therefore, the outcomes do not necessarily depend on our phenomenological method. Additionally, it is hoped that the polarization of background microwave photons would allow for the measurement of the gravity wave contribution to the anisotropy of the CMBR, providing new tests for cosmological models. The work aims to develop a significant tool for studying different dark energy models comparatively using the features of gravitational wave evolution. The work is structured as follows: Background equations of gravitational waves in a flat FRW universe are studied in section II. The dynamics of the gravitational waves in the background of different holographic dark energy models are explored in section III. Finally, the paper ends with a brief conclusion in section IV.

\section{Background Equations of Gravitational Waves in Flat FRW Universe}
In this section, we discuss the background equations of gravitational waves in flat FRW spacetime. The standard flat FRW metric is given by
\begin{equation}\label{2.1}
ds^2=-dt^2+a^2(t)\left(dx^2+dy^2+dz^2\right),
\end{equation}
where $a(t)$ is the cosmological scale factor. We can obtain the Einstein’s field equations as
\begin{equation}\label{2.2}
\frac{\dot a^2}{a^2}=\frac{1}{3}\left(\rho_m+\rho_D\right),
\end{equation}
\begin{equation}\label{2.3}
\frac{2\ddot{a}}{a}+\frac{\dot a^2}{a^2}=-\left(p_m+p_D\right),
\end{equation}
where $\rho_m$, $\rho_D$ are the energy density of matter and dark energy respectively. Moreover $p_m$ and $p_D$ are the pressure of matter and dark energy respectively.

The continuity equations are given by
\begin{equation}\label{2.4}
\dot\rho_m+\frac{3\dot a}{a}\left(\rho_m+p_m\right)=0,
\end{equation}
and
\begin{equation}\label{2.5}
\dot\rho_D+\frac{3\dot a}{a}\left(\rho_D+p_D\right)=0.
\end{equation}
Considering that matter satisfies the equation of state given by $p_m=\omega_m\rho_m$, (where $\omega_m$ is the constant equation of state (EoS) parameter), we obtain from the continuity equation
\begin{equation}\label{2.6}
\rho_m=\rho_{m_0}a^{-3\left(1+\omega_m\right)}=\rho_{m_0}\left(1+z\right)^{3\left(1+\omega_m\right)},
\end{equation}
where $\rho_{m_0}$ is the present value of matter density and $z=\frac{a_0}{a}-1$ is the cosmological redshift.

The governing equations of the evolution of gravitational waves for a flat universe are as follows:
\begin{equation}\label{2.7}
    \ddot{\eta}(t)-\frac{\dot{a}}{a}\dot{\eta}
(t)+\left(\frac{\xi^2}{a^2}-2\frac{\ddot{a}}{a} \right)\eta(t)=0,
\end{equation}
which can be written as:
\begin{equation}\label{2.8}
    \eta^{''}(z)-\left(\frac{\ddot{a}}{\dot{a}^2}-\frac{2}{a} \right)a^2 \eta^{'}(z)
    +\frac{a^4}{\ddot{a}^2}\left(\frac{\xi^2}{a^2}-2\frac{\ddot{a}}{a} \right)\eta(z)=0,
\end{equation}
where $'\equiv \frac{d}{dz}$.

Next we express the equations (\ref{2.2}) and (\ref{2.3}) as follows:
\begin{equation}\label{2.9}
\frac{\dot a^2}{a^2}=H_0^2 X(z)~~~ \mbox{and}~~~\frac{2\ddot{a}}{a}=-H_0^2 Y(z),
\end{equation}
where
\begin{equation}\label{2.10}
    X(z)=\frac{1}{3H_0^2}\left(\rho_m+\rho_D \right)
\end{equation}
and 
\begin{equation}\label{2.11}
    Y(z)=\frac{1}{3H_0^2}\left[\rho_m \left(1+3\omega_m \right)+\rho_D\left(1+3\omega_D\right)\right],
\end{equation}
where $\omega_m=\frac{p_m}{\rho_m},~ \omega_D=\frac{p_D}{\rho_D}$ are the equation of state parameter of matter and dark energy respectively. $H_0$ is the present value of the Hubble parameter $H=\frac{\dot a}{a}$
Exploiting equations (\ref{2.9}),(\ref{2.10}) and (\ref{2.11}), equation (\ref{2.8}) reads:
\begin{equation}\label{2.12}
    \eta^{''}(z)+\frac{1}{2(1+z)}\left( \frac{Y(z)}{X(z)}+4\right) \eta^{'}(z)
    +\frac{4}{Y^2(z)H_0^4}\left(\xi^2+\frac{H_0^2 Y(z)}{(1+z)^2}\right)\eta(z)=0,
\end{equation}
 The characteristics of the gravitational waves for various holographic dark energy models in a flat FRW Universe are going to be investigated in the following sections.

\section{Gravitational Waves for Holographic Dark Energy Model}
There is a fundamental notion that the holographic dark energy density ought to be correlated with the inverse squared Infrared cutoff $L$, specifically,
\begin{equation}
\rho_{de}=\frac{3c}{\kappa^{2}L^{2}}
\end{equation}
where $\kappa^2$ is the gravitational constant and $c$ is a parameter. It is widely acknowledged as long as one accepts the holographic principle's cosmological application; nevertheless, there isn't a clear consensus on what the appropriate infrared cutoff should be. In the subsequent subsections we will study the evolution of gravitational waves for various holographic dark energy models.

\subsection{Model 1: Ricci Holographic dark energy}
The Ricci scalar curvature for a flat FRW spacetime is given by
\begin{equation}\label{1.1}
R=-6\left(\dot{H}+2H^2\right)
\end{equation}
We consider a dark energy component that, according to the holographic principle, is proportional to the inverse of the squared Ricci scalar curvature radius and originates from unidentified physics. The energy density of the Ricci Holographic dark energy (RHDE) model is given by \cite{rhde1},
\begin{equation}\label{1.2}
\rho_{Ricci}=-3\left(\alpha' R\right)=3\alpha\left(\dot{H}+2H^2\right)
\end{equation}
where $c$ and other constants have been absorbed in $\alpha$.

Here we assume a power-law form of scale factor:
\begin{equation}\label{1.3}
a(t)=b_0 t^n,
\end{equation}
where $b_0, n$ are constants.\\
Correspondingly $\rho_{Ricci}$ takes the form
\begin{equation}\label{1.4}
\rho_{Ricci}=3\alpha b_0^{\frac{2}{n}} n (2n-1) (1+z)^{\frac{2}{n}}
\end{equation}
Hubble parameter for Ricci HDE model is obtained as
\begin{equation}\label{1.5}
H(z)=H_0 \left[\Omega_{m0} (1+z)^{3(1+\omega_m)}+ \frac{n\alpha(2n-1)b_0^{\frac{2}{n}}}{H_0^2} (1+z)^{\frac{2}{n}}\right]^{\frac{1}{2}},
\end{equation}
where $\Omega_{m0}=\frac{\rho_{m0}}{3 H_0^2}$.\\

\begin{figure}[hbt!]
\centering
\includegraphics[height=2.5in]{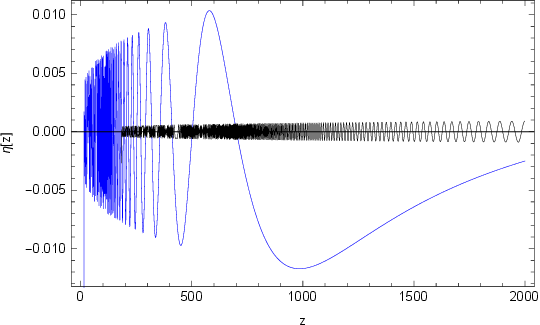}
\centering \caption{Plot of $\eta(z)$ against redshift $z$, for Ricci HDE with the parameters $H_0=70$, $\omega_m=0.0001$, $n=5$,$\alpha=0.5$,$b_0=0.002$ and $\Omega_{m0}=0.25$. The parameter $\xi=5\times10^{11}$ for the black line and $\xi=2\times10^9$ for blue line.}
\label{Fig1}
\end{figure}

The differential equation of gravitational wave for Ricci HDE model can be obtained from eqn.(\ref {2.12}), where $X(z), Y(z)$ are given in the following form: 
\begin{equation}\label{1.6}
X(z)= \Omega_{m0} (1+z)^{3(1+\omega_m)}+ \frac{n\alpha(2n-1)b_0^{\frac{2}{n}}}{ H_0^2} (1+z)^{\frac{2}{n}},
\end{equation}

\begin{equation}\label{1.7}
Y(z)=(1+3\omega_m)\Omega_{m0} (1+z)^{3(1+\omega_m)}+\frac{2\alpha(2n-1)(1-n)b_0^{\frac{2}{n}}}{H_0^2}  (1+z)^{\frac{2}{n}}.
\end{equation}
The solution of the differential equation is shown in the fig.(\ref{Fig1}). From the figure we clearly see the wave-like trajectory of $\eta(z)$. There is a significant dependence of the amplitude of the gravitational waves on the parameter $\xi$. For greater value of $\xi$ we have waves of smaller amplitude and vice versa. Moreover with the evolution of the universe the gravitational wave amplitude gradually decrease indicating a gradual decay of the waves as they move away from the source. We also see that for Ricci HDE the waves are concentrated (shorter time period) around the present time ($z=0$) but they become stretched (longer time period) as we move back in time ($z\rightarrow \infty$). This show that the spacetime ripples become smaller but crowded with the evolution of the universe. This can be a direct consequence of the increasing entropy of the universe.\\

\subsection{Model 2: Tsallis Holographic dark energy}
The energy density of the Tsallis holographic dark energy (THDE) is given by,
\begin{equation}\label{2.1a}
\rho_{thde}=\beta L^{2\delta-4}
\end{equation}
where $\beta$ and $\delta$ are unknown parameters. Here we assume the Hubble horizon as the IR cutoff, i.e. $L=H^{-1}$ and get the energy density of THDE as \cite{thde1},
\begin{equation}\label{2.2a}
\rho_{thde}=\beta H^{4-2\delta}
\end{equation}

$\rho_{thde}$ for power law form of scale factor reads
\begin{equation}\label{2.3a}
\rho_{thde}=\beta n^{4-2\delta} b_0^{\frac{4-2\delta}{n}} (1+z)^{\frac{4-2\delta}{n}}
\end{equation}
Hubble parameter for THDE model is obtained as
\begin{equation}\label{2.4a}
H(z)=H_0 \left[\Omega_{m0} (1+z)^{3(1+\omega_m)}+ \beta n^{4-2\delta}  b_0^{\frac{4-2\delta}{n}} (1+z)^{\frac{4-2\delta}{n}}\right]^{\frac{1}{2}},
\end{equation}
where $\Omega_{m0}=\frac{\rho_{m0}}{3 H_0^2}$.\\

The differential equation of gravitational wave for THDE model can be obtained from eqn.(\ref {2.12}), where $X(z), Y(z)$ are given in the following form:   
\begin{equation}\label{2.5a}
X(z)= \Omega_{m0} (1+z)^{3(1+\omega_m)}+ \frac{\beta  b_0^{\frac{4-2\delta}{n}} n^{4-2\delta}}{3 H_0^2} (1+z)^{\frac{4-2\delta}{n}}
\end{equation}

\begin{equation}\label{2.6a}
Y(z)=(1+3\omega_m)\Omega_{m0} (1+z)^{3(1+\omega_m)}+\frac{2\beta(2-\delta-n)  b_0^{\frac{4-2\delta}{n}} n^{4-2\delta}}{3n H_0^2} (1+z)^{\frac{4-2\delta}{n}}.
\end{equation}

\begin{figure}[hbt!]
\centering
\includegraphics[height=2.5in]{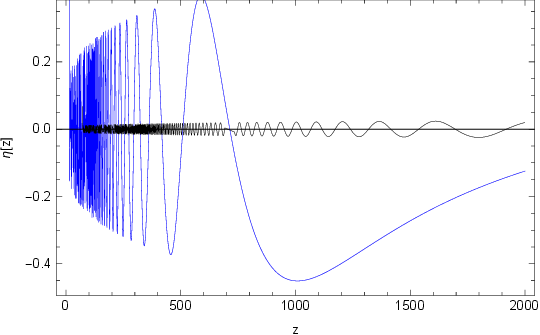}
\centering \caption{Plot of $\eta(z)$ against redshift $z$, for Tsallis HDE  with the parameters $H_0=70$, $\omega_m=0.0001$, $n=2$,$\delta=2$, $\beta=2.3$,$b_0=0.002$ and $\Omega_{m0}=0.24$. The parameter $\xi=5\times10^{11}$ for the black line and $\xi=2\times10^9$ for the blue line.}
\label{Fig2}
\end{figure}

The solution of the differential equation is shown in the fig.(\ref{Fig2}). In the case of THDE, we see that the gravitational waves of larger amplitude are produced compared to the case of Ricci HDE for smaller values of the parameter $\xi$ (blue line). For higher values of $\xi$ the nature of the waves in this case is more stretched. This is basically due to the presence of the parameter $\delta$ in the THDE energy density. Moreover, for the THDE model, the system will be less dynamic as compared to the Ricci HDE model due to the absence of the time gradient of the Hubble parameter. This gives a comparative idea about the two HDE models and their effects on the evolution of the gravitational waves. \\

\subsection{Model 3: Renyi Holographic dark energy}
The energy density of Renyi Holographic dark energy (RenHDE) is given by
\begin{equation}\label{3.1}
\rho_{rhde}=\frac{3c^2}{L^2}\left(1+\pi \delta L^2\right)^{-1}
\end{equation}\label{3.2}
where $c$ and $\delta$ are parameters. Here also we consider the Hubble horizon as the IR cut-off and get the energy density as \cite{renhde1},
\begin{equation}\label{3.3}
\rho_{rhde}=3c^2\frac{H^2}{\left(1+\pi \delta/ H^2\right)}
\end{equation}

$\rho_{rhde}$ for power law form of scale factor becomes
\begin{equation}\label{3.4}
\rho_{rhde}=\frac{3c^2 n^4 b_0^\frac{4}{n} (1+z)^\frac{4}{n}}{\left(\pi \delta +n^2 b_0^\frac{2}{n} (1+z)^\frac{2}{n}\right)}
\end{equation}

Hubble parameter for RHDE model is computed as
\begin{equation}\label{3.5}
H(z)=H_0 \left[\Omega_{m0} (1+z)^{3(1+\omega_m)}+\frac{3c^2 n^4 b_0^\frac{4}{n} (1+z)^\frac{4}{n}}{\left(\pi \delta +n^2 b_0^\frac{2}{n}(1+z)^\frac{2}{n}\right)}\right]^\frac{1}{2},
\end{equation}

The differential equation of gravitational wave for RHDE model can be obtained from eqn.(\ref {2.12}), where $X(z), Y(z)$ are given in the following form: 
\begin{equation}\label{3.6}
X(z)= \Omega_{m0} (1+z)^{3(1+\omega_m)}+  \frac{3c^2 n^4 b_0^\frac{4}{n} (1+z)^\frac{4}{n}}{H_0^2\left(\pi \delta +n^2 b_0^\frac{2}{n} (1+z)^\frac{2}{n}\right)}
\end{equation}

\begin{equation}\label{3.7}
Y(z)=(1+3\omega_m)\Omega_{m0} (1+z)^{3(1+\omega_m)}+\frac{2 c^2 n^3 b_0^\frac{4}{n} (1+z)^\frac{4}{n} \Big( n^2 (1-n)b_0^\frac{2}{n}(1+z)^\frac{2}{n}+2 \pi \delta(2-n) \Big)}{H_0^2\left(\pi \delta +n^2 b_0^\frac{2}{n} (1+z)^\frac{2}{n}\right)^2}
\end{equation}

\begin{figure}[hbt!]
\centering
\includegraphics[height=2.5in]{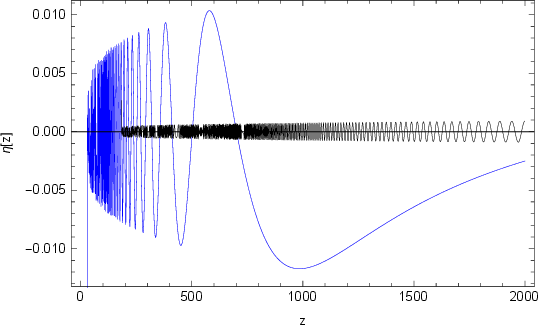}
\centering \caption{Plot of $\eta(z)$ against redshift $z$ for Renyi HDE with the parameters $H_0=70$, $\omega_m=0.0001$, $n=5$,$\delta=1.8$,$c=250$, $b_0=20$ and $\Omega_{m0}=0.25$.. The parameter $\xi=5\times10^{11}$ for the black line and $\xi=2\times10^9$ for the blue line.}
\label{Fig3}
\end{figure}

The solution of the differential equation for RHDE has been plotted in Fig. (\ref{Fig3}). From the plot, we see that the characteristics of the gravitational waves closely mimic those of the Ricci HDE. The amplitude of the waves is smaller in this case compared to the case of THDE.

\subsection{Model 4: Barrow Holographic dark energy}
In the Barrow Holographic dark energy (BHDE) model, the energy density is given by {arxiv 2203.14464}.
\begin{equation}\label{4.1}
\rho_{bhde}= C L^{\Delta-2}
\end{equation}
where $C$ is the parameter whose dimension is given by $L^{-2-\Delta}$ and $R_{h}$ represents the future event horizen. If $\Delta=0$ then Eq.(\ref{4.1}) takes the form $\rho_{de}= C L^{-2}$, where $C=c^2 {M_{p}}^2$, $M_p$ beig the Planck mass. 

\vspace{5mm}

\subsubsection{BHDE with Hubble Horizon Cut-off}
If we consider the Hubble horizon cut-off ($L=H^{-1}$) we get the energy density as,
\begin{equation}\label{4.2}
\rho_{bhde_h}= C H^{2-\Delta}
\end{equation}

$Rho_{bhde_h}$ for power law form of scale factor reads
\begin{equation}\label{4.3}
\rho_{bhde_h}=C n^{2-\Delta} b_0^{\frac{2-\Delta}{n}} (1+z)^{\frac{2-\Delta}{n}}
\end{equation}

Hubble parameter for BHDE model with Hubble horizon cut-off is obtained as
\begin{equation}\label{4.4}
H(z)=H_0 \left[\Omega_{m0} (1+z)^{3(1+\omega_m)}+\frac{C n^{2-\Delta}b_0^{\frac{2-\Delta}{n}}}{3 H_0^2} (1+z)^{\frac{2-\Delta}{n}}\right]^{\frac{1}{2}},
\end{equation}
where $\Omega_{m0}=\frac{\rho_{m0}}{3 H_0^2}$.\\

The differential equation of gravitational wave for BHDE model with Hubble horizon cut-off can be obtained from eqn.(\ref {2.12}), where $X(z), Y(z)$ are given in the following form:
\begin{equation}\label{4.5}
X(z)= \Omega_{m0} (1+z)^{3(1+\omega_m)}+ \frac{C n^{2-\Delta}b_0^{\frac{2-\Delta}{n}}}{3 H_0^2} (1+z)^{\frac{2-\Delta}{n}}
\end{equation}

\begin{equation}\label{4.6}
Y(z)=(1+3\omega_m)\Omega_{m0} (1+z)^{3(1+\omega_m)}+\frac{C(2-\Delta-2n) n^{1-\Delta}b_0^{\frac{2-\Delta}{n}}}{3 H_0^2} (1+z)^{\frac{2-\Delta}{n}}.
\end{equation}

\begin{figure}[hbt!]
\centering
\includegraphics[height=2.5in]{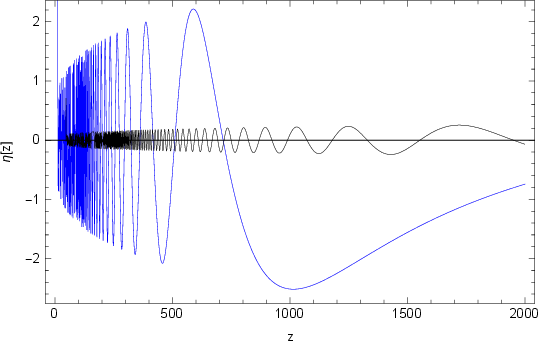}
\centering \caption{Plot of $\eta(z)$ against redshift $z$ for Barrow HDE withe Hubble horizon cut-off. The parameter $\xi=5\times10^{11}$ for the black line and $\xi=2\times10^9$ for the blue line.The other parameters are taken as $H_0=70$, $\omega_m=0.001$, $n=5$, $\Delta=3.2$,$b_0=250$, $C_1=2$ and $\Omega_{m0}=0.25$.}
\label{Fig1a}
\end{figure}

\subsubsection{BHDE with Granda-Oliveros (GO) cut-off}
However, the Granda-Oliveros (GO) cut-off is very widely used in the case of BHDE where the IR cut-off is given by,
\begin{equation}\label{4.7}
L^{-2}=(\alpha H^2+\beta \dot{H})
\end{equation}
where $\alpha$ and $\beta$ are parameters. Implementing Eq.(\ref{4.7}) in Eq.(\ref{4.1}) we get energy density to be of the form \cite{bhde1}
\begin{equation}\label{4.8}
\rho_{bhde_{go}}= 3 {M_{p}}^2 (\alpha_1 H^2+\beta_1 \dot{H})^{1-\frac{\Delta}{2}}
\end{equation}
Exploiting the power law form of scale factor the BHDE model with GO cut-off reads
\begin{equation}\label{4.9}
\rho_{bhde_{go}}= 3 {M_{p}}^2 (\alpha_1 n^2-\beta_1 n)^{1-\frac{\Delta}{2}} b_0^{\frac{2-\Delta}{n}} (1+z)^{\frac{2-\Delta}{n}}
\end{equation}
Hubble parameter for the BHDE model with Granda-Oliveros (GO) cut-off is obtained as
\begin{equation}\label{4.10}
H(z)=H_0 \left[\Omega_{m0} (1+z)^{3(1+\omega_m)}+\frac{{M_{p}}^2 (\alpha_1 n^2-\beta_1 n)^{1-\frac{\Delta}{2}}b_0^{\frac{2-\Delta}{n}}}{H_0^2} (1+z)^{\frac{2-\Delta}{n}}\right]^{\frac{1}{2}},
\end{equation}
where $\Omega_{m0}=\frac{\rho_{m0}}{3 H_0^2}$.\\

The differential equation of gravitational wave for the BHDE model with GO cut-off can be obtained from eqn.(\ref {2.12}), where $X(z), Y(z)$ are given in the following form:
\begin{equation}\label{4.11}
X(z)= \Omega_{m0} (1+z)^{3(1+\omega_m)}+\frac{{M_{p}}^2 b_0^{\frac{2-\Delta}{n}}(\alpha_1 n^2-\beta_1 n)^{1-\frac{\Delta}{2}}}{H_0^2} (1+z)^{\frac{2-\Delta}{n}}
\end{equation}

\begin{equation}\label{4.12}
Y(z)=(1+3\omega_m)\Omega_{m0} (1+z)^{3(1+\omega_m)}+\frac{{M_{p}}^2(\Delta-2-2n)(\alpha_1 n^2-\beta_1 n)^{1-\frac{\Delta}{2}}b_0^{\frac{2-\Delta}{n}} }{n H_0^2} (1+z)^{\frac{2-\Delta}{n}}.
\end{equation}

\begin{figure}[hbt!]
\centering
\includegraphics[height=2.5in]{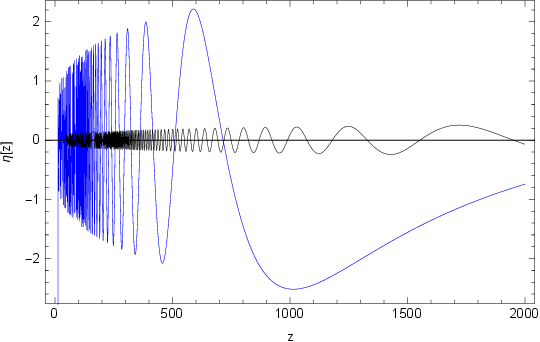}
\centering \caption{Plot of $\eta(z)$ against redshift $z$ for Barrow HDE with GO cut-off. The parameter $\xi=5\times10^{11}$ for the black line and $\xi=2\times10^9$ for the blue line. The other parameters are taken as $H_0=70$, $\omega_m=0.001$, $n=5$, $\Delta=3.2$,$b_0=250$, $C_1=2$ and $\Omega_{m0}=0.25$, $\alpha_1=2$, $\beta_1=5$, $M_P=2.176434 \times 10^{-8}$.}
\label{Fig1a}
\end{figure}


In Figs.(4) and (5) we have obtained the plots for the gravitational wave perturbations against the redshift $z$ in the case of BHDE for the Hubble-horizon cut-off and the GO cut-off respectively. The plots exhibit marked similarity between the two. This probably indicates that there is minimum dependence of the evolution of gravitational waves on the cut-off scale of the BHDE. This is a very important result derived in this study.

\subsection{Model 5: Sharma-Mittal Holographic dark energy}
The energy density for the Sharma-Mittal Holographic dark energy (SMHDE) density for the Hubble horizon cut-off is given by \cite{smhde1},
\begin{equation}\label{5.1}
\rho_{smhde}=\frac{C^2H^4}{d}\left[\left(1+\frac{\delta\pi}{H^2}\right)^{d/\delta}-1\right]
\end{equation}
where $d$ and $\delta$ are two parameters.
Applying the power law form of scale factor the SMHDE model takes the form
\begin{equation}\label{5.2}
\rho_{SMHDE}= \frac{C^2 n^4 b_0^{\frac{4}{n}}}{d}  (1+z)^{\frac{4}{n}} \left[ \left(1+\frac{\delta \pi}{n^2 b_0^\frac{2}{n}} (1+z)^\frac{-2}{n}\right)^{\frac{d}{\delta}}-1\right]
\end{equation}
Hubble parameter for SMHDE model cut-off is obtained as
\begin{equation}\label{5.3}
H(z)=H_0 \left[\Omega_{m0} (1+z)^{3(1+\omega_m)}+\frac{C^2 n^4 b_0^{\frac{4}{n}}}{3 d H_0^2}  (1+z)^{\frac{4}{n}} \left( \left(1+\frac{\delta \pi}{n^2 b_0^\frac{2}{n}} (1+z)^\frac{-2}{n}\right)^{\frac{d}{\delta}}-1\right)\right]^{\frac{1}{2}},
\end{equation}
where $\Omega_{m0}=\frac{\rho_{m0}}{3 H_0^2}$.\\

The differential equation of gravitational wave for SMHDE model can be obtained from eqn.(\ref {2.12}), where $X(z), Y(z)$ are given in the following form:
\begin{equation}\label{5.4}
X(z)=\left[\Omega_{m0} (1+z)^{3(1+\omega_m)}+\frac{C^2 n^4 b_0^{\frac{4}{n}}}{3 d H_0^2}  (1+z)^{\frac{4}{n}} \left( \left(1+\frac{\delta \pi}{n^2 b_0^\frac{2}{n}} (1+z)^\frac{-2}{n}\right)^{\frac{d}{\delta}}-1\right)\right]
\end{equation}

\begin{eqnarray*}\label{5.5}
Y(z)=(1+3\omega_m)\Omega_{m0} (1+z)^{3(1+\omega_m)}+\frac{2 C^2 n^4 b_0^{\frac{4}{n}}}{3 d H_0^2}  (1+z)^{\frac{4}{n}} \left( \left(1+\frac{\delta \pi}{n^2 b_0^\frac{2}{n}} (1+z)^\frac{-2}{n}\right)^{\frac{d}{\delta}}-1\right)
\end{eqnarray*}
\begin{equation}
  \left[-1+\frac{d(1+z)^{\frac{-2}{n}}}{n^3 b_0^{\frac{2}{n}}}\left( \frac{2n^2 b_0^{\frac{2}{n}}(1+z)^{\frac{2}{n}}}{d}-\frac{\pi\left(1+\frac{\delta \pi}{n^2 b_0^\frac{2}{n}} (1+z)^\frac{-2}{n}\right)^{\frac{d}{\delta}-1}}{\left(1+\frac{\delta \pi}{n^2 b_0^\frac{2}{n}} (1+z)^\frac{-2}{n}\right)^{\frac{d}{\delta}}-1}\right) \right]  
\end{equation}

\begin{figure}[hbt!]
\centering
\includegraphics[height=2.5in]{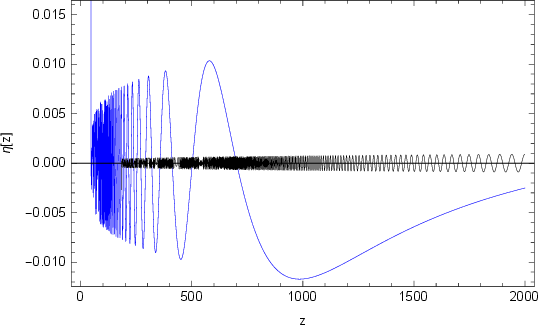}
\centering \caption{Plot of $\eta(z)$ against redshift $z$ for Sharma Mittal HDE. The parameter $\xi=5\times10^{11}$ for the black line and $\xi=2\times10^9$ for the blue line.The other parameters are taken as $H_0=70$, $\omega_m=0.0001$, $n=5$, $\Delta=3.2$,$b_0=20$, $C=250$, $d=200$, $\delta=1.8$ and $\Omega_{m0}=0.25$.}
\label{Fig1a}
\end{figure}

In Fig.(6) we have generated the plot for the gravitational wave parameter $\eta$ against the redshift $z$ for SMHDE. This plot is quite different from the other plots because its amplitude is much smaller. Moreover, the ripples are very much concentrated around $z=0$ (present time) compared to the early universe. There is also a decrease in the amplitude as the wave evolves from the early to the late universe due to the stronger effect of dark energy.

\subsection{Model 6: Kaniadakis Holographic dark energy}
The energy density for the Kaniadakis Holographic dark energy (KHDE) density for the Hubble horizon cut-off is given by \cite{khde1},
\begin{equation}
\rho_{khde}=\frac{C^2H^4}{\kappa}sinh\left(\frac{\kappa\pi}{H^2}\right)
\end{equation}
where $C$ and $\kappa$ are two parameters.

Exploiting the power law form of scale factor the KHDE model reads:

\begin{equation}\label{6.1}
\rho_{KHDE}=\frac{c^2 n^4 b_0^{\frac{4}{n}}(1+z)^{\frac{4}{n}}}{\kappa} \sinh \left( \frac{\kappa \pi}{n^4 b_0^{\frac{2}{n}}(1+z)^{\frac{2}{n}}}\right)
\end{equation}

The hubble parameter for the KHDE model is computed as follows:
\begin{equation}\label{6.2}
    H=H_0\left[\Omega_{m0} (1+z)^{3(1+\omega_m)}+ \frac{c^2 n^4 b_0^{\frac{4}{n}}(1+z)^{\frac{4}{n}}}{3\kappa H_0^2} \sinh \left( \frac{\kappa \pi}{n^4 b_0^{\frac{2}{n}}(1+z)^{\frac{2}{n}}}\right)\right]^{\frac{1}{2}},
\end{equation}
where $\Omega_{m0}=\frac{\rho_{m0}}{3 H_0^2}$.\\

The differential equation of gravitational wave for KHDE model can be obtained from eqn.(\ref {2.12}), where $X(z), Y(z)$ are given in the following form:

\begin{equation}\label{6.3}
    X(z)=\Omega_{m0} (1+z)^{3(1+\omega_m)}+ \frac{c^2 n^4 b_0^{\frac{4}{n}}(1+z)^{\frac{4}{n}}}{3\kappa H_0^2} \sinh \left( \frac{\kappa \pi}{n^4 b_0^{\frac{2}{n}}(1+z)^{\frac{2}{n}}}\right)
\end{equation}
and
\begin{equation}\label{6.4}
    Y(z)=(1+3\omega_{m})\Omega_{m0}(1+z)^{3(1+\omega_m)}+ \frac{2c^2 n^4 b_0^{\frac{4}{n}}(1+z)^{\frac{4}{n}}}{3\kappa H_0^2} \sinh \left( \frac{\kappa \pi}{n^4 b_0^{\frac{2}{n}}(1+z)^{\frac{2}{n}}}\right)\left(  -1+\frac{2}{n}-\frac{\kappa \pi \tanh\left( \frac{\kappa \pi}{n^4 b_0^{\frac{2}{n}}(1+z)^{\frac{2}{n}}}\right)}{n^3 b_0^{\frac{2}{n}}(1+z)^{\frac{2}{n}} }\right)
\end{equation}
\begin{figure}[hbt!]
\centering
\includegraphics[height=2.5in]{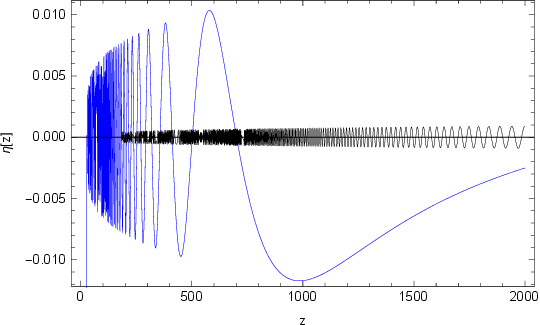}
\centering \caption{Plot of $\eta(z)$ against redshift $z$ for KHDE. The parameter $\xi=5\times10^{11}$ for the black line and $\xi=2\times10^9$ for the blue line.The other parameters are taken as $H_0=70$, $\omega_m=0.0001$, $n=5$, $\kappa=0.2$,$C=250$, $b_0=20$ and $\Omega_{m0}=0.25$.}
\label{Fig1a}
\end{figure}

In Fig.(7) we have generated the plot for the perturbation parameter $\eta$ against the redshift $z$. Here similarity is seen with the case of RHDE. The amplitude and other characteristics of the waves seem to be similar to those obtained for RHDE.

\section{Conclusion}
In this work, we have investigated the evolution of gravitational waves in the presence of different holographic dark energy models. Six different models of holographic dark energy are considered and the evolution of gravity waves is investigated. The basic idea is to explore the effects of black hole thermodynamics in the evolution of gravity waves. Initially, we considered the flat Friedmann-Robertson-Walker (FRW) spacetime. We examined energy densities for both dark energy and matter by examining distinct conservation equations for each component. We developed the perturbation equations regulating gravitational wave evolution with respect to redshift $z$ in the FRW Universe backdrop using the field equations. The properties of gravitational waves for HDE models were then discussed. Because the gravitational wave differential equations in this model are complicated, we used graphical analysis to derive wave trajectories at different redshift ranges. The amplitudes diminish with time as the universe evolves, $z\rightarrow0$, as is evident in all the figures. This is an expected result because it is known that dark energy dynamics do tend to diminish the amplitude of the spacetime ripples \cite{rip}. Now the reason for this is that as the universe evolves dark energy domination is much more pronounced. So with greater effect coming from the dark sector, the amplitudes of the waves diminish. So in our work, we have obtained reliable results having support from the literature. Our work also helps to qualitatively distinguish between the different holographic dark energy models. Since the dark energy sector has minimum or no interaction with other sectors except gravitational interaction, it becomes very difficult to study their qualitative or quantitative features by standard methods. Herein lies the importance of the method proposed in this work involving gravitational waves. Since the gravitational wave is a purely gravitational phenomenon they have wide applications in dark energy dynamics. This is the basic idea behind our work. Moreover, the imprint of black hole thermodynamics in our work differentiates it from any other dark energy models and makes the work more interesting. Although we did not find exact analytical solutions for the models considered, but the methodology described in this work is quite significant and can be used to compare different dark energy models. As far as exact analytical solutions are concerned it is completely model-dependent and may be obtainable for some other models which would be excellent. But it should be admitted that there is a lot of room for improvement in this work, like comparison with observational data, and the anisotropy spectrum analysis of the cosmic microwave background radiation. The absence of an analytical solution hindered us in some way. But as discussed earlier this is simply a restriction imposed by the models and not a defect in the methodology. These improvements will be addressed in future projects.

\section*{Acknowledgments}

PR acknowledges the Inter-University Centre for Astronomy and
Astrophysics (IUCAA), Pune, India for granting visiting
associateship.

\section*{Data Availability}

No new data were generated in this paper.

\section*{Conflict of Interests}

There are no conflicts of interest in this paper.

\section*{Funding Statement}

No funding was received for this paper.


\end{document}